\documentclass[twocolumn,aps,amssymb,floatfix,superscriptaddress,preprintnumbers]{revtex4}

\usepackage{graphicx}
\usepackage{bm}
\usepackage{amsmath}
\usepackage{amssymb}
\usepackage{amsfonts}
\usepackage{float}
\usepackage{hyperref}
\usepackage{dsfont}  
\usepackage{slashed} 
\usepackage{booktabs}
\usepackage{multirow}
\usepackage{subfigure}
\usepackage[sort&compress]{natbib}

\newcommand{\be}{\begin{equation}}  
\newcommand{\ee}{\end{equation}}  
\newcommand{\beq}{\begin{eqnarray}}  
\newcommand{\eeq}{\end{eqnarray}}

\newcommand{\bea}{\begin{eqnarray}}
\newcommand{\eea}{\end{eqnarray}}

\newcommand{\MSb}{{\overline{\rm MS}}}

\begin{document}
\title{Transversity parton distribution functions from lattice QCD}
\author{Constantia Alexandrou}
\affiliation{
Department of Physics,
  University of Cyprus,
  P.O. Box 20537,
  1678 Nicosia,
  Cyprus}
\affiliation{
  Computation-based Science and Technology Research Center,
  The Cyprus Institute,
  20 Kavafi Street,
  Nicosia 2121,
  Cyprus}
\author{Krzysztof Cichy}
\affiliation{Faculty of Physics, Adam Mickiewicz University, Umultowska 85, 61-614 Pozna\'{n}, Poland}
\author{Martha Constantinou}
\affiliation{Department of Physics,  Temple University,  Philadelphia,  PA 19122 - 1801,  USA}
\author{Karl Jansen}
\affiliation{NIC, DESY,
  Platanenallee 6,
  D-15738 Zeuthen,
  Germany}
\author{Aurora Scapellato}
\affiliation{
Department of Physics,
  University of Cyprus,
  P.O. Box 20537,
  1678 Nicosia,
  Cyprus}
\affiliation{University of Wuppertal, Gau\ss str. 20, 42119 Wuppertal, Germany}
\author{Fernanda Steffens}
\affiliation{Institut f\"ur Strahlen- und Kernphysik, Rheinische
  Friedrich-Wilhelms-Universit\"at Bonn, Nussallee 14-16, 53115 Bonn}

\preprint{DESY 18-104}

\begin{abstract}

\noindent 

We present the first direct calculation of the transversity parton distribution function within the nucleon from lattice QCD. The calculation is performed using simulations with the light 
quark mass fixed to its physical value and at one value of the lattice spacing. Novel elements of the calculations are non-perturbative 
renormalization and extraction of a formula for the matching to light-cone PDFs. Final results are presented in the $\MSb$ scheme at a 
scale of $\sqrt{2}$ GeV.

\end{abstract}

\maketitle 
\bibliographystyle{apsrev}

\noindent\textit{Introduction:} 
The proton is the simplest baryon. The strong force which is governing the internal dynamics of baryons makes them immensely complicated systems even already for the proton as the simplest baryon, with several non-trivial structure functions that are actively being investigated both theoretically and experimentally.
One of the most important tools to understand QCD dynamics in a hadron are parton distribution functions (PDFs), which give
information on the hadron spin and momentum distribution among its constituent quarks and gluons.
During the past four decades detailed information on PDFs have been accumulated as function of the longitudinal momentum fraction
$x$ of the nucleon momentum that they carry \cite{Aidala:2012mv,Accardi:2016ndt,Gao:2017yyd}.
These encompass the class of chiral-even distributions:
$f_1(x)$, for unpolarized partons; and $g_1(x)$, for helicity PDF in a longitudinally polarized nucleon. The remaining collinear
transversity PDF, $h_1(x)$, which is defined as the difference between probabilities to find
a parton spin aligned and anti-aligned to the transversely
polarized parent nucleon, is considerably less known, because it is chiral-odd  \cite{Jaffe:1991kp}.
This requires $h_1(x)$ to couple to another chiral-odd function in order to be accessed by
experiments, excluding the traditional totally inclusive processes that have successfully unveiled
different facets of the inner structure of the nucleon.

The transverse momentum dependent (TMD) Collins fragmentation function (FF) $H_1^\perp(x,p_\perp)$ \cite{Collins:1992kk}
is chiral-odd and couples to the TMD transversity $h_1(x,k_\perp^2)$ in the
Collins single-spin asymmetry, which also contains information on the spin averaged TMD FF 
and unpolarized TMD PDF. 
The FFs can be
obtained independently from $e^+e^-$
annihilation into dihadrons of transverse momentum $p_\perp$, with data available
from BELLE \cite{Abe:2005zx,Seidl:2008xc}, BABAR \cite{Garzia:2012za}, and BESIII \cite{Garzia:2012za} collaborations.
In conjunction with HERMES \cite{Airapetian:2010ds} and COMPASS \cite{Alekseev:2008aa,Adolph:2014zba}, semi-inclusive
deep-inelastic scattering (SIDIS) TMD data for single hadron
production, $h_1(x,k_\perp^2)$, are then extracted. An ansatz \cite{Rogers:2015sqa} is used to disentangle
the dependence on the momentum fraction from the transverse momentum on both the TMD FFs and the TMD PDFs to, finally, obtain
the $h_1(x)$ distribution.
Alternatively, in order to avoid the use of TMDs and to benefit from collinear factorization, it has
been advocated \cite{Collins:1993kq,Jaffe:1997hf,Bacchetta:2011ip} to use dihadron SIDIS cross section
data to relate the measured asymmetry directly to $h_1(x)$.
This asymmetry, integrated over the transverse momentum of the hadrons, is proportional to a simple product of $h_1(x)$ and
an integral of the chiral-odd dihadron FF. The disadvantage of this route, as compared to the former case that
uses single hadron production, is that the available data are less precise. In addition, collinear factorization is
problematic at large $x$ \cite{Moffat:2017sha}.
Nevertheless, both procedures have been recently employed in global analyses for the extraction of the $x$-dependence of the nucleon transversity \cite{Anselmino:2007fs,Anselmino:2015sxa,Radici:2015mwa,Kang:2015msa,Lin:2017stx,Radici:2018iag}.

It is clear from this
discussion that the extraction of the transversity distributions is sensitive to the FFs, with or without TMDs, and also to the TMD PDFs. Accordingly, 
reliable and independent information on $h_1(x)$ is important in its own right, as the computation of the $x$-dependence of PDFs from first principles 
is a milestone in itself. In addition to that, it has the potential to constrain the FFs, TMDs FFs, and TMDs PDFs. It is also clear from the Monte Carlo 
analysis of global data of Ref.~\cite{Lin:2017stx} that SIDIS data by themselves do not impose tight constraints either to the $x$-dependence of the 
distributions or to the tensor charges associated with them (see, e.g., SIDIS data in Fig.~\ref{fig:transv_PDFs}). The isovector tensor charge, $g_T {=} \int_0^1(h_1^u(x) - h_1^d(x))dx $, in particular, is 
found to be 0.9(8) when only SIDIS data are used. The inclusion of lattice data for $g_T$ in the global fit, on the other hand, moves $g_T$ to 1.0(1), 
while reducing the uncertainty in the corresponding $x$-dependence of the distributions by a factor of at least 3, showing the enormous impact that 
information from first principles has on the determination of the transversity distributions. The analysis of Ref.~\cite{Radici:2018iag}, which uses dihadron 
interference FFs in SIDIS for a single-fit analysis of electron-proton and proton-proton data, however, has found that the individual $u$ and 
$d$ charges are incompatible between their results and those of Ref.~\cite{Lin:2017stx}. 
A further comparison between the charges obtained from phenomenology and the ones from lattice QCD shows additional tension~\cite{Ye:2016prn}, 
reinforcing the idea of a transverse spin puzzle. Given that the tensor charge is a candidate to constrain physics beyond the Standard 
Model~\cite{Dubbers:2011ns,DelNobile:2013sia,Bhattacharya:2011qm,Courtoy:2015haa}, it follows that a direct computation of the $x$-dependence of $h_1(x)$ 
has impact in different branches of particle and nuclear physics from imposing limits on the hadron momentum fraction $x$ and the transverse momentum 
$p_\perp$-dependence of the FFs, to constraints on the extensions of the Standard Model itself. The overarching importance of transversity is translated into the inclusion 
of experiments aimed at measuring transversity in the JLab 12 GeV program~\cite{Ye:2016prn,Dudek:2012vr}, and at the future Electron Ion 
Collider~\cite{Accardi:2012qut,Aschenauer:2014twa}, which will explore the  large and small $x$ regions of $h_1(x)$, respectively.

As mentioned above, the PDFs extracted from experimental data use input from theoretically motivated parameterizations. However, it would be imperative to have 
prediction from first principles, and an ideal formulation is lattice QCD. In this letter, we present the first {\it ab initio} computation of the $x$-dependence for the 
isovector transversity PDF in lattice QCD, as described below.

\vspace*{0.5cm}
\noindent{\it Quasi-PDFs:} Light-cone PDFs cannot be directly evaluated on a Euclidean lattice, but certain information is obtained from their moments in lattice QCD~\cite{Constantinou:2014tga,Constantinou:2015agp,Alexandrou:2015yqa,Alexandrou:2015xts,Syritsyn:2014saa}. The reconstruction of light-cone PDFs from their moments, however, is a difficult task, due to increasing statistical noise for high moments, as well as due to a power divergence 
for moments with more than three covariant derivatives. Recently, a new direction has been proposed by X.\ Ji~\cite{Ji:2013dva} to obtain the Bjorken-$x$ dependence of PDFs using lattice QCD, the so-called quasi-PDFs, defined for the transversity case as
\begin{equation}
\label{eq:quasi_pdf}
\widetilde{h}_1(x,\Lambda,P){=}\hspace*{-0.1cm}\int_{-\infty}^{+\infty}\hspace*{-0.1cm}\frac{dz}{4\pi}\,
e^{-ixP_3z}\,M_{h_1}(P,z)\,.
\end{equation}
$\Lambda{\sim} 1/a$ is a UV cut-off and $M_{h_1}(P,z)$ is the non-local matrix element
\begin{eqnarray}
\label{eq:def_ME}
M_{h_1}(P,z) &=& \langle N(P)\,\vert \,\mathcal{O}(z)\,\vert N(P)\rangle\,, \\[0.5ex]
\mathcal{O}(z) &=& \overline{\psi}(z)\frac{\sigma_{31}+\sigma_{32}}{2} W(z;0)\psi(0)\,,
\label{eq:def_O}
\end{eqnarray} 
between two nucleon states $\vert N\rangle$ having a spatial momentum $\vec{P}{=}(0,0,P_3)$, if the boost is in the $z$-direction. The Wilson line is taken 
along the direction of the momentum and has a length varying from zero up to the half of the spatial extent of the lattice. It connects a quark ($\psi$) and 
antiquark ($\overline{\psi}$) with a Dirac matrix $\sigma_{3i}$ in the transversity case, where the second index denotes the spatial direction of the nucleon spin, perpendicular to the direction of the momentum.
For this particular choice of the Dirac structure, $M_{h_1}(P,z)$ does not exhibit mixing with other operators~\cite{Constantinou:2017sej}. 
Note that each of the term of Eq.~(\ref{eq:def_O}) is projected appropriately to avoid statistical noise contamination.
To get from quasi-PDFs to light-cone PDFs, bare matrix elements are renormalized, and then matching and target mass corrections (TMCs) are used, provided that 
nucleon momenta are large enough for Large Momentum Effective Theory (LaMET) to be applicable~\cite{Ji:2013dva,Ji:2014gla}.
State-of-the-art results on the spin-averaged and helicity PDFs using the same setup appear in Ref.~\cite{Alexandrou:2018pbm} 
(see also~\cite{Chen:2018xof}).

\vspace*{0.5cm}
\noindent {\it Numerical setup}: We use one  ensemble of gauge configurations, produced with the Iwasaki improved gauge action~\cite{Iwasaki:2011np,Abdel-Rehim:2013yaa} 
and two degenerate ($N_f{=}2$) twisted mass clover-improved fermions~\cite{Frezzotti:2003ni,Sheikholeslami:1985ij} with their masses fixed approximately to their physical 
value~\cite{Abdel-Rehim:2015pwa}. The ensemble has a lattice volume of $48^3 \times 96$, lattice spacing $a\simeq 0.093$~fm, resulting in a spatial extent of $L\simeq 4.5$~fm, 
and a pion mass $m_\pi\simeq 130$~MeV.

In this work, we focus on the isovector flavor combination, $h_1^{u{-}d}$, that receives only connected contributions. We also obtain $h_1^{\bar{d}}(x){-}h_1^{\bar{u}}(x)$ given the crossing relation $h_1^{\bar{q}}(x){=}\,{-}h_1^{q}(-x)$~\cite{Chang:2014jba}. Lattice data are extracted from a ratio of $M_{h_1}(P,z)$ projected at zero momentum transfer, over the corresponding two-point function. 

A state with the quantum numbers of the nucleon is created at a source ($t{=}0$), annihilated at the sink ($T_{\rm sink}$), and couples with $\mathcal{O}$ at time-slice $t_{\rm ins}$. For sufficiently large time separations, $T_{\rm sink}$, the desired matrix element of the nucleon ground state are then extracted. To improve the overlap with the ground state, we use Gaussian smeared interpolating fields~\cite{Gusken:1989ad,Alexandrou:1992ti} with APE-smeared gauge links~\cite{Albanese:1987ds}. The three-point functions require the computation of an all-to-all propagator from all positions of the sink to all spatial insertion points, and is evaluated by performing sequential inversions through the sink~\cite{Martinelli:1988rr}.
Moreover, to increase the overlap of the nucleon interpolating field with the ground state of the boosted particle, we use the momentum smearing technique~\cite{Bali:2016lva}, which has also been applied in our previous quasi-PDFs calculations~\cite{Alexandrou:2016jqi,Alexandrou:2018pbm}. For each value of the nucleon momentum used in this work, we tune the parameter entering the momentum smearing function in order to maximize the overlap with the boosted nucleon. This results in a significant reduction of the statistical uncertainties on the nucleon correlators. Quasi-PDFs have been computed for three values of the nucleon momentum, namely $P{=}\frac{6\pi}{L},\,\frac{8\pi}{L},\,\frac{10\pi}{L}$ ($\sim$ 0.83, 1.11, 1.38 GeV) with $T_{\rm sink}$ set to $12a\simeq 1.12$~fm. To keep the statistical errors approximately the same for the different boosts, we increase the number of configurations, as can be seen in Table~\ref{tab:statistics}.
\begin{table}[h!]
\centering
{\small
\renewcommand{\arraystretch}{1.2}
\renewcommand{\tabcolsep}{6pt}
\vspace*{-0.2cm}
  \begin{tabular}{ l| c | c | c }
   & $\frac{6\pi}{L}$ (0.83GeV) & $\frac{8\pi}{L}$ (1.11GeV)& $\frac{10\pi}{L}$ (1.38GeV) \\
    \hline\hline
    $N_{conf}$ & 100 & 425 & 811 \\ \hline
    $N_{meas.}$ & 9600 & 38250 & 72990  \\
    \hline
  \end{tabular}
  }
  \caption{Statistics used in this calculation for each momentum. $N_{conf}$ is the number of configurations and $N_{meas.}$ the total number of measurements.}
  \label{tab:statistics}
  \vspace*{-0.2cm}
  \end{table}
For each configuration, we use multiple random source positions and for $P{=}\frac{8\pi}{L},\,\frac{10\pi}{L}$ we employ the {truncated solver method~\cite{Bali:2009hu} to reduce the
computational cost. 
For each source position, we average over correlation functions where the momentum and Wilson line are aligned along all possible directions (namely, $\pm x$, $\pm y$, $\pm z$), 
to increase the number of measurements. To reduce statistical uncertainties in the renormalized $M_{h_1}(P,z)$ we apply five steps of three-dimensional stout smearing~\cite{Morningstar:2003gk} 
to the gauge links entering the operator $\mathcal{O}$. We find no dependence of the renormalized $M_{h_1}(P,z)$ on the number of iterations as discussed in Ref.~\cite{Alexandrou:2018pbm}. In Fig.~\ref{fig:bare_ME}, we show the momentum dependence of the bare $M_{h_1}(P,z)$. As expected, for higher momentum boosts, they decay to zero faster with increasing $z$ and the imaginary part becomes more prominent.
\begin{figure}[h]
\hspace*{-27.5cm}
\includegraphics[scale=0.8]{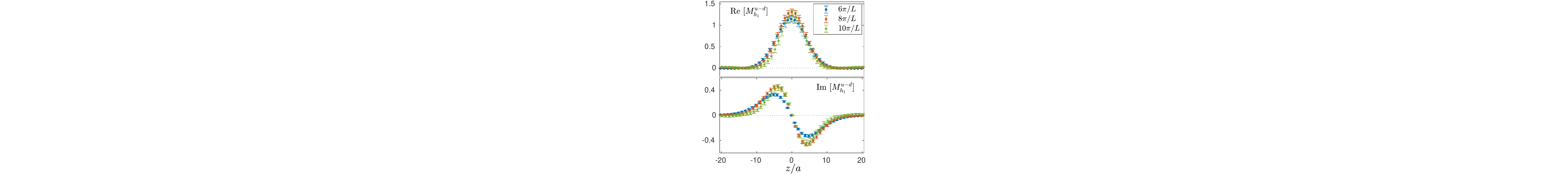}
\caption{Real (upper) and imaginary part (lower) of the bare $M_{h_1}(P,z)$ at 5 stout steps, for $P{=}\frac{6\pi}{L}$ (blue), $P{=}\frac{8\pi}{L}$ (red) and $P{=}\frac{10\pi}{L}$ (green), as a function of the Wilson line length, $z/a$.}
\label{fig:bare_ME}
\end{figure}

We examine ground state dominance on each momentum as we vary $T_{\rm sink}$. We use three values for $T_{\rm sink}$ namely, $8 a,\,10 a,\,12 a$. Convergence is seen between $T_{\rm sink}{=}10a$ and $T_{\rm sink}{=}12a$ and therefore we conclude that ground-state dominance is achieved for $T_{\rm sink}{=}12a$. In particular, the real part of Eq.~(\ref{eq:def_ME}) is compatible between $T_{\rm sink}{=}10a$ and $T_{\rm sink}{=}12a$, while the imaginary part in the small-$z$ region is more influenced by excited states.
Note however, that boosting the nucleon with momenta higher than what we presently use, is expected to require even larger values of 
$T_{\rm sink}$ to suppress excited states. For the momenta we employ, we have also checked the dispersion relation as a test of cut-off effects, and no deviations from the continuum 
energy-momentum dependence are observed.

 \vspace*{0.5cm}
\noindent\textit{Towards light-cone PDFs}:
The ME of Eq.~(\ref{eq:def_ME}) have the additional complication of a power-like divergence of the form $|z|/a$ inherited by the Wilson line. The latter, together with logarithmic 
divergences, must be eliminated prior to taking the Fourier transform of Eq.~(\ref{eq:quasi_pdf}). 
Recently, we proposed a non-perturbative RI$'$-type prescription~\cite{Alexandrou:2017huk} that eliminates all divergences in a consistent and reliable fashion (see also \cite{Green:2017xeu} for an alternative prescription). The renormalization procedure is outlined in Ref.~\cite{Alexandrou:2017huk} and we use the same setup as in 
Ref.~\cite{Alexandrou:2018pbm}. Using the results of Ref.~\cite{Constantinou:2017sej} we convert the renormalization functions to the $\overline{\rm MS}$ scheme and evolve to $\mu{=}\sqrt{2}$ GeV, to match the scale used in the phenomenological data~\cite{Lin:2017stx}. A linear extrapolation ($(a\,\mu_0)^2 {\to} 0$) is applied to eliminate residual dependence on the RI$'$ scale $\mu_0$.

As a last step, we extract the renormalized light-cone PDF, $h_1(x,\mu)$, from renormalized quasi-PDF, $\widetilde{h}_1\left(x,\mu,P_3\right)$, via a matching
procedure \cite{Xiong:2013bka,Alexandrou:2015rja,Chen:2016fxx,Wang:2017qyg,Stewart:2017tvs,Izubuchi:2018srq}:
\begin{eqnarray}
\label{eq:matching}
h_1( x,\mu )
  =\! \int_{-\infty}^\infty \frac{d\xi}{|\xi|}\, \delta C \left(
   \xi, \frac{\xi \mu}{xP_3} \right) \widetilde{h}_1 \left(\frac{x}{\xi},\mu,P_3\right)\!,
\end{eqnarray}
where $\delta C(\xi,\frac{\xi\mu}{x P_3})$ is the matching kernel. $\delta C$ can be computed perturbatively within QCD, because the infrared (IR) physics is the same for quasi- and light-cone PDFs. $\delta C$ has been computed at one-loop order for the transversity case~\cite{Xiong:2013bka}, using a hard transverse momentum cut-off and a quark mass to regularize the UV and the IR divergences, respectively. Here we perform a computation using a gluon mass to regularize the IR divergences, and DR to regularize the UV divergences. The resulting kernel is independent of the gluon mass and takes quasi-PDFs renormalized in the $\MSb$ scheme to light-cone PDFs in this scheme. It reads:

\begin{widetext}
\begin{eqnarray}
\label{eq:kernel}
\delta C\left( \xi, \frac{\xi \mu}{xP_3} \right)=\delta(1-\xi)+\frac{\alpha_s}{2\pi}C_F\left\{
\begin{array}{ll}
\displaystyle \left[\frac{2 \xi}{1-\xi}\ln\frac{\xi}{\xi-1} + \frac{2}{\xi}\right]_+
&\, \xi>1,
\\[10pt]
\displaystyle \left[\frac{2\xi}{1-\xi}\left(
\ln\frac{4 x^2 P_3^2}{\xi^2\mu^2}+ln(\xi(1-\xi))\right) - \frac{2 \xi }{1-\xi}  \right]_+
&\, 0<\xi<1,
\\[10pt]
\displaystyle \left[-\frac{2\xi}{1-\xi}\ln\frac{\xi}{\xi-1} + \frac{2}{1-\xi}\right]_+
&\, \xi<0\,.
\end{array}\right.
\end{eqnarray}
\end{widetext}
The plus prescriptions in Eq.~(\ref{eq:kernel}) are all at $\xi{=}1$.
The contributions outside the physical region in Eq.~(\ref{eq:kernel}) have their origin
exclusively in the one-loop correction to quasi-PDF. Conversely, quasi-PDFs are UV finite inside the physical region. Thus,
unlike light-cone PDF, the UV divergence in quasi-PDF appears when integrating the momentum fraction in the one-loop wave-function correction to $\pm\infty$. These divergences, which behave as  $-2/\xi$ ($\xi{>}1$)and $-2/(1{-}\xi)$ ($\xi{<}0$), have been subtracted in Eq.~(\ref{eq:kernel}). From the Ward identity, the integrated one-loop vertex correction is renormalized by the same terms. This ensures that the norm of the distributions is automatically preserved by the matching, i.e.\ $\int_{-\infty}^{\infty} dx\, h_1(x,\mu) {=} \int_{-\infty}^{\infty} dx\, \widetilde{h}_1(x,\mu,P_3)$, and
$\int_{-\infty}^{\infty} d\xi\, \delta C(\xi,\frac{\xi\mu}{x P_3}){=}1$ that holds for every value of $P_3$. Because particle number conservation is built inside the matching, the finite limits of integration imposed by the lattice data also conserve the norm.

\vspace*{0.5cm}
\noindent\textit{Final Results:} 
A combination of the renormalization, matching procedure and application of the TMCs allow the reconstruction of light-cone PDFs, which we present in this section. 
In Fig.~\ref{fig:quasi_evol.pdf}, we show the effect of each step of this procedure for $P{=}\frac{10\pi}{L}$, i.e.\ we start with the renormalized quasi-PDF ($\widetilde{h}_1^{u-d}$), apply matching (${h'_1}^{u-d}$) and finally include TMCs ($h_1^{u-d}$), which leads to the final estimate of the transversity PDF. As can be seen, application of the matching shifts the peak of the distribution towards $x{=}0$ and increases it, as expected. We also find that TMCs are small, but non-negligible, and mostly affect the small-$x$ region.
\begin{figure}[h]
\hspace*{-27cm}
\includegraphics[scale=0.77]{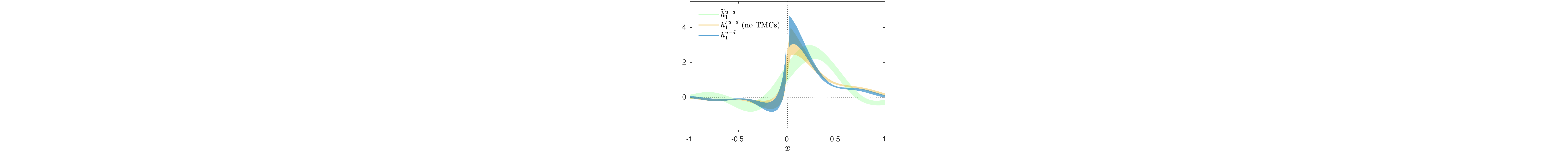}
\caption{Renormalized quasi-PDF, $\widetilde{h}_1^{u-d}$ (green), PDF after matching,  ${h'_1}^{u-d}$ (orange) and after TMCs, $h_1^{u-d}$ (blue), as a function of Bjorken-$x$ for $P{=}\frac{10\pi}{L}$.}
\label{fig:quasi_evol.pdf}
\end{figure}

Our final results are shown in Fig.~\ref{fig:transv_PDFs} at a scale of $\sqrt{2}$ GeV ($h_1^{u-d,{\rm lattice}}$). For clarity we only show $P{=}\frac{10\pi}{L}$, as the dependence on the 
nucleon momentum is small for most regions of $x$. We find that for the large and positive $x$ region, the data at momentum 
$P{=}\frac{10\pi}{L}$ have milder oscillatory behavior, an effect that originates from the use of finite momentum. As can bee seen in the plot, $h_1^{u-d,{\rm lattice}}$ in the large negative-$x$ 
nicely approach zero. For demonstration purposes, we include in the same plot phenomenological fits on SIDIS data~\cite{Lin:2017stx}, as well as SIDIS data constrained using lattice 
estimates of $g_T$ (``SIDIS+lattice'')~\cite{Lin:2017stx}. The statistical uncertainties of the lattice PDFs are strikingly smaller than the phenomenological fits of the SIDIS data. This also holds for the 
``SIDIS+lattice'' data that have much smaller uncertainties than the unconstrained SIDIS values. The comparison favors the direct extraction of the transversity PDF using the quasi-PDFs 
method, in terms of uncertainties and reliability in the extraction. Using the data at $P{=}\frac{10\pi}{L}$, we obtain $g_T{=}1.10(34)$ 
by integrating over $x$ within the interval $[-1,\,1]$. This value can be compared with the renormalized $M_{h_1}(P,0)$ that gives a value $g_T{=}1.09(11)$, which is compatible with the aforementioned integration and with the extraction in Ref.~\cite{Alexandrou:2017qyt}.

\begin{figure}[h]
\hspace*{-27cm}
\includegraphics[scale=0.77]{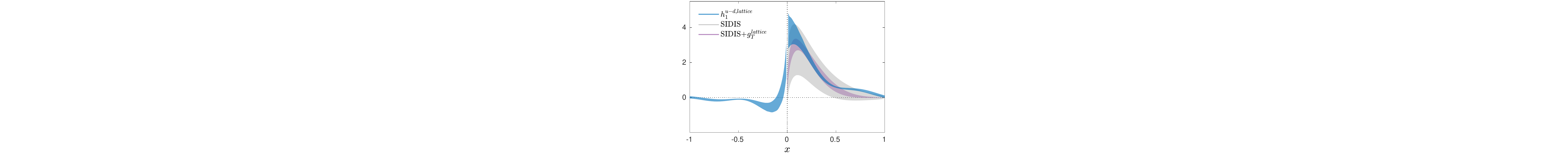}
\caption{Transversity PDF for $P{=}\frac{10\pi}{L}$ (blue) as a function of  Bjorken-$x$. The phenomenological fits have been obtained using SIDIS data (grey)~\cite{Lin:2017stx} and SIDIS data constrained using $g_T^{\rm lattice}$ (purple)~\cite{Lin:2017stx}.}
\label{fig:transv_PDFs}
\end{figure}

\noindent\textit{Summary and Prospects}:
This paper presents a state-of-the-art direct calculation of the transversity PDF for the isovector flavor combination. The novelty of this work is the improvement of the computation in all fronts, that is, simulations at physical quark masses~\cite{Abdel-Rehim:2015pwa,Abdel-Rehim:2015owa}, employment of a non-perturbative renormalization 
program~\cite{Alexandrou:2017huk}, and application of a cut-off independent and renormalized matching between quasi-PDFs and light-cone PDFs; the latter was developed in this work.

A number of careful investigations have been performed to study systematic uncertainties. We find that excited states are suppressed for a source-sink separation of 1.12 fm and nucleon momentum up to 1.4 GeV. The twisted mass formulation has the advantage of automatic ${\cal O}(a)$-improvement, and calculation of the dispersion relation indicates small cut-off effects. Another investigation is the extraction of the tensor charge found to be $g_T{=}1.10(34)$, which is compatible with the dedicated study of Ref.~\cite{Alexandrou:2017qyt}, that is $g_T{=}1.06(1)$ at $\mu{=}\sqrt{2}$ GeV. In addition, the renormalized $M_{h_1}(P,z)$ is completely independent of the stout smearing, which demonstrates the success of the renormalization program.

The  great progress in the \textit{ab initio} calculation of PDFs is paving the way for further improvements. New computer architectures can lead to the extraction of reliable estimates for PDFs using higher nucleon momenta, that would presently require computational resources beyond what is currently available to eliminate excited states contamination. Development of new techniques to tackle the exponential increase of the statistical noise with 
respect to the source-sink separation and higher momenta will be crucial for the quasi-PDFs calculations.  If successful, future calculations at higher momenta will allow one to investigate higher-twist effects, while the continuum 
limit can be reliably taken using simulations with at least three values of the lattice spacing. Another direction that could potentially be improved is the calculation of renormalization functions, by 
utilizing the perturbative calculation of Ref.~\cite{Constantinou:2017sej}, and a technique to eliminate finite lattice spacing effects to order ${\cal O}(g^2a^\infty)$. The latter has successfully 
improved the renormalization functions of local operators~\cite{Alexandrou:2015sea}. 

The {\textit{ab initio} calculation of the transversity PDF is a powerful demonstration of the predictability of lattice QCD. The final lattice results presented in Fig.~\ref{fig:transv_PDFs} 
compare favorably to the phenomenological fits that carry very large uncertainties preventing any meaningful conclusions. Even though constraining the SIDIS data with lattice estimates on the 
tensor charge reduces the errors significantly, the overall uncertainties are still very large~\cite{Lin:2017stx,Radici:2018iag}. This reflects the large experimental difficulties to extract transversity PDF, which, unlike the chiral-even distributions, can only be accessed if additional information from FFs is given. Thus, a computation from first principles is imperative, even more  when we compare it to its unpolarized and helicity counterparts, which can be directly probed in totally inclusive experiments. All the above point to the direction that a prediction of the transversity 
PDF directly from lattice QCD is a major breakthrough that opens new directions in understanding the immensely rich nucleon structure.

\vspace*{0.3cm}
\begin{acknowledgements}

We express gratitude to all members of ETMC for the pleasant collaboration. 
M.C. would like to thank Zein-Eddine Meziani for many fruitful and inspiring discussions.
F.S thanks Nobuo Sato for helpful discussions and for providing the SIDIS phenomenological
parameterizations. This work has received funding from the European Union's Horizon 2020 
research and innovation programme under the Marie Sk\l{}odowska-Curie grant agreement 
No 642069 (HPC-LEAP). K.C.\ is supported by the National Science Centre grant SONATA 
BIS no.\ 2016/22/E/ST2/00013. F.S.\ is funded by the Deutsche Forschungsgemeinschaft (DFG) project number 392578569.
M.C. acknowledges financial support by the U.S. Department of Energy, Office of Nuclear Physics, within
the framework of the TMD Topical Collaboration, as well as, by the National Science Foundation
under Grant No.\ PHY-1714407. 
This research used resources of the Oak Ridge Leadership Computing Facility, which is a DOE Office of Science User 
Facility supported under Contract DE-AC05-00OR22725, Prometheus supercomputer 
at the Academic Computing Centre Cyfronet AGH in Cracow (grant ID \textit{quasipdfs}), Okeanos 
supercomputer at the Interdisciplinary Centre for Mathematical and Computational Modelling in 
Warsaw (grant IDs gb70-17, ga71-22), Eagle supercomputer at the Poznan Supercomputing and Networking Center (grant no.\ 346). 
\end{acknowledgements}

\bibliography{references_transv}

\end{document}